\documentclass[12pt]{report}
\usepackage[english]{babel}
\usepackage{subfigure}
\usepackage{graphicx}
\usepackage{epsfig}
\usepackage{amsmath}
\usepackage{amssymb}
\usepackage{amsfonts}
\usepackage{slashed}
\usepackage{bbm}
\usepackage{cancel}
\usepackage{wrapfig}
\usepackage[font=scriptsize,labelfont=bf]{caption}
\usepackage[section]{placeins}
\usepackage{fullpage}

\begin{document}

\begin{titlepage}

\centerline{\huge The CAPTAIN Detector and Physics Program}

\vspace{0.3in}

\centerline{\large \today}

\vspace{0.3in}


\medskip


H. Berns$^4$,
H. Chen$^3$,
D. Cline$^{6}$,
J. Danielson$^{10}$,
Z. Djurcic$^{2}$,
S. Elliott$^{10}$,
G. Garvey$^{10}$,
V. Gehman$^{9}$,
C. Grant$^4$,
E. Guardincerri$^{10}$,
R. Kadel$^{9}$,
T. Kutter$^{11}$,
D. Lee$^{10}$,
K. Lee$^{6}$,
Q. Liu$^{10}$,
W. Louis$^{10}$,
C. Mauger$^{10}$,
C. McGrew$^{12}$,
R. McTaggart$^{14}$,
J. Medina$^{10}$,
W. Metcalf$^{11}$,
G. Mills$^{10}$,
J. Mirabal-Martinez$^{10}$,
S. Mufson$^{8}$,
E. Pantic$^4$,
O. ProkofievS. Mufson$^{7}$,
V. Radeka$^3$,
J. Ramsey$^{10}$,
K. Rielage$^{10}$,
H. Sahoo$^{2}$,
C. Sinnis$^{10}$,
M. Smy$^{5}$,
W. Sondheim$^{10}$,
I. Stancu$^{1}$,
R. Svoboda$^4$,
M. Szydagis$^4$,
C. Taylor$^{10}$,
A. Teymourian$^{6}$,
C. Thorn$^3$,
C. Tull$^{9}$,
M. Tzanov$^{11}$,
R. Van de Water$^{10}$,
H. Wang$^{6}$,
C. Yanagisawa$^{12}$,
A. Yarritu$^{10}$, 
C. Zhang$^{13}$

\centerline{\it $^{1}$University of Alabama, Tuscaloosa, AL 35487-0324, USA}
\centerline{\it $^{2}$Argonne National Laboratory, Argonne, IL 60473, USA}
\centerline{\it $^{3}$Brookhaven National Laboratory, Upton, NY 11973-5000, USA}
\centerline{\it $^{4}$University of California at Davis, Davis, CA 95616, USA}
\centerline{\it $^{5}$University of California at Irvine, Irvine, CA 92697-4575, USA}
\centerline{\it $^{6}$University of California at Los Angeles, Los Angeles, CA 90095-1547, USA}
\centerline{\it $^{7}$Fermi National Accelerator Laboratory, Batavia, IL 60510, USA}
\centerline{\it $^{8}$Indiana University, Bloomington, IN 47405, USA}
\centerline{\it $^{9}$Lawrence Berkeley National Laboratory, Berkeley, CA 94720-8153, USA}
\centerline{\it $^{10}$Los Alamos National Laboratory, Los Alamos, NM 87545, USA}
\centerline{\it $^{11}$Louisiana State University, Baton Rouge, LA 70803-4001}
\centerline{\it $^{12}$State University of New York at Stony Brook, Stony Brook, NY 11798}
\centerline{\it $^{13}$University of South Dakota, Vermillion, SD 57069, USA}
\centerline{\it $^{14}$South Dakota State University, Brookings, SD 57007, USA}

\medskip

\vspace{0.25in}

\end{titlepage}

\newpage

\tableofcontents

\newpage

\begin{abstract}

The Cryogenic Apparatus for Precision Tests of Argon Interactions with Neutrino (CAPTAIN) 
program is designed to make measurements of scientific importance to long-baseline neutrino 
physics and physics topics that will be explored by large underground detectors.  
CAPTAIN began as part of a Los Alamos National Laboratory (LANL) Laboratory Directed Research and 
Development (LDRD) project and has evolved into a multi-institutional collaboration.  
The program employs two detectors.  
The CAPTAIN detector is a liquid argon TPC deployed in a portable and evacuable 
cryostat that can hold a total of 7700 liters of liquid argon.  Five tons of liquid argon are instrumented 
with a 2,000 channel liquid argon TPC and a photon detection system.  The cryostat has ports that can hold optical windows 
for laser calibration and windows for the introduction of charged particle beams.  
The materials for the detector are currently being acquired.  Assembly is anticipated to begin August of 
2013, with a commissioning period ending in the summer of 2014.  During commssioning, laser calibration 
and cosmic-ray data will be taken and analyzed.  Subsequent to the commissioning phase, the detector will 
be moved to a high-energy neutron beamline that is part of the Los Alamos Neutron Science Center.  
The neutron data will be used to measure cross-sections of spallation products that are backgrounds to 
measurements of neutrinos from a supernova burst and cross-sections of events that mimic the electron 
neutrino appearance signal in long-baseline neutrino physics.
The data will also be used to develop strategies for counting neutrons and evaluating their energies 
in a liquid argon TPC that are important for the total neutrino energy measurement in the analysis of long-
baseline neutrinos.
The prototype detector is being fabricated in a cryostat supplied by the University of California at 
Los Angeles (UCLA) group and consists of a 1,000 channel liquid argon time-projection chamber (TPC).  
Fabrication also begins in August of 2013, but will be completed more quickly than the CAPTAIN detector.  
The prototype detector will allow an end-to-end test of all components in time to make adjustments to the 
scheme employed by CAPTAIN.  
The prototype will collect cosmic-ray and laser calibration data earlier than will be possible in CAPTAIN 
allowing the development of analysis techniques at an earlier date.  
Finally, the prototype will allow for testing of calibration and other ideas in parallel to the running of 
CAPTAIN.

Subsequent to the neutron running, the CAPTAIN detector will be moved to a neutrino source.
There are several possible neutrino sources of interest.  
The two most likely neutrino possibilities are an on-axis run in the NuMI (Neutrinos at the Main Injector) 
beamline at Fermi National Accelerator Laboratory (FNAL) and a run in the neutrino source produced by the 
Spallation Neutron Source (SNS) at Oak Ridge National Laboratory.  
An on-axis run at NUMI produces more than one million events of interest in a two or three year run at 
neutrino energies between 1 and 10 GeV.  
The neutrino studies are complementary to the MicroBooNE experiment, which will measure similar 
interactions at a lower, complementary energy range - 0.5 to 2 GeV.  
Many important exclusive and inclusive charged and neutral current channels can be measured 
by such a run.  
In addition, a detailed evaluation of strategies to measure the total neutrino energy including neutrons 
emitted by the argon nucleus are made possible by this run.
The SNS produces a neutrino source as a byproduct of its neutron production.  
The neutrinos result from the decays stopped positively charged pions and muons.  
The neutrino energy spectrum produced from these decays is a broad spectrum up to 50 MeV.
If located close to the spallation target, CAPTAIN can detect several thousand events per year in 
the same neutrino energy regime where neutrinos from a supernova burst are.  
Measurements at the SNS yield a first measurement of the cross-section of neutrinos on argon 
in this important energy regime.  
In addition, this would be the first measurement of low-energy neutrinos in a liquid argon TPC.
This is critical for the interpretation of a supernova burst in the LBNE far detector and will 
greatly affect the specifications for the light collection and DAQ systems.

\end{abstract}

\newpage

\chapter{Introduction}

\indent

The development of the CAPTAIN (Cryogenic Apparatus for Precision Tests of 
Argon Interactions with Neutrinos) experiment began as a part of an Los Alamos National Laboratory (LANL) Laboratory Directed Research and Development (LDRD) project.  
The three-year LDRD project is designed to enhance the science associated with the 
Long-Baseline Neutrino Experiment (LBNE).  The project involves scientists from three groups 
in two divisions engaged in an integrated effort of theory, simulation and detector work.
CAPTAIN is 
the major hardware effort of the LDRD project and involves significant contributions from the 
entire experimental and technical staff supported by the project.

From the early stages of development, the LANL team received significant technical assistance from Brookhaven National Laboratory (BNL), Fermi National Accelerator Laboratory (FNAL), and the University of California at Los Angeles (UCLA).  In additional to technical assistance, the UCLA group has provided a cryostat that will contain the CAPTAIN prototype detector.

The scientific focus of the CAPTAIN program is to make measurements important for the development of the Long-Baseline Neutrino Experiment (LBNE).  LBNE is a broad scientific program being developed in the United States as an international partnership. It consists of an intense neutrino beam produced at FNAL, a highly capable set of neutrino detectors on the FNAL campus, and a large underground liquid argon time projection chamber (TPC) at Sanford Underground Research Facility (SURF), giving a 1300 km oscillation baseline.   The high-intensity neutrino beam will allow high precision measurements of neutrino and anti-neutrino mixing separately. enabling detailed studies of neutrino oscillations, including measurements of the mass hierarchy, CP violation,  and non-standard neutrino interactions (NSI).
In addition to serving as a far detector for the long-baseline neutrino physics program, the large underground far detector enables a broad scientific program that includes searches for nucleon decay mediated by beyond the standard model physics, the study of neutrinos from galactic supernova bursts, indirect searches for the annihilation products of dark matter particles and the detailed study of atmospheric neutrinos.  
The CAPTAIN program impacts several of the topics that make up the LBNE physics program via two prongs of study:  low energy neutrino physics and medium energy neutrino physics.  

The primary low-energy neutrino signal in LBNE will be from charged-current (CC) electron neutrinos.  
\begin{equation}
\nu_e +   ^{40}Ar   \rightarrow   e^{-}  +  ^{40}K^{*}
\end{equation}
The transition to the ground state of potassium is forbidden, so in general, the CC electron neutrino signal is accompanied by one or more de-excitation gamma-rays.  
The cross-section has never been measured and has theoretical uncertainties in the range of 10 to 15 percent.  
Backgrounds to detecting this process in the LBNE far detector are from cosmic ray muon induced spallation processes.  
In addition, the final states consist of a single electron and a ``halo'' of gammas, making triggering and analysis challenging.
The three most important spallation processes are the photonuclear interaction between the muon and an argon nucleus, neutron induced spallation of the argon nucleus and charged pion spallation of the argon nucleus.  In the latter two processes, the neutrons and pions are produced by muon interactions in the rock surrounding the detector or in the detector itself.  
CAPTAIN will provide several important inputs for LBNE.  First, CAPTAIN will measure neutron 
spallation processes in a high-energy neutron beam at the Los Alamos Neutron Science Center 
(LANSCE).  Next, CAPTAIN will measure CC and neutral current (NC) low-energy neutrino cross-sections with a stopped-pion neutrino source at the Spallation Neutrino Source (SNS) at Oak Ridge National Laboratory (ORNL).  The measurements will be the first low-energy neutrino measurements in a liquid argon TPC, and will provide practical inputs to the design of the LBNE far detector light collection and DAQ system.

Medium-energy neutrino interactions are poorly understood on any nucleus.  
There is a dearth of neutrino-argon data in the 1 to 10 GeV neutrino energy regime.  
LBNE will use data in this energy regime to make high-precision measurements of neutrino 
oscillation phenomena.  Irrespective of the source of medium-energy neutrinos - beam or 
atmospheric -, neutrino oscillation studies depend on having well-constrained determinations 
of three quantities for each neutrino event:  the neutrino flavor, the distance from the point of production, and the neutrino energy.  In general, CC neutrino interactions in this energy regime will result in a lepton, the emission of several hadrons and a residual nucleus.  While charged 
hadrons will be well-identified and measured, neutrons are harder to measure.  They travel 
some distance from the neutrino interaction vertex and deposit energy via elastic and inelastic collisions with argon nuclei.  The determination of the neutrino energy is therefore a significant challenge and possible systematic limitation to the LBNE neutrino physics program.  CAPTAIN will address issues associated with neutrino energy reconstruction.  First, CAPTAIN will make a detailed study of neutron interactions with argon as a function of neutron energy up to neutron kinetic energies of 800 MeV.  Using these data, the collaboration will develop methodologies to constrain the neutron energy in a neutrino interaction.  Next, with an on-axis NuMI run, where the neutrino beam is measured with a variety of detectors, CAPTAIN will make a detailed study of neutrino interactions on argon.  The methodologies developed with the neutron data will be put to the test with the high statistics neutrino data.  In addition, CAPTAIN will contain a large 
fraction of the hadronic component of the interactions, so the data will be an important test-bed for developing automated liquid argon reconstruction techniques critical for LBNE.

In the next chapter, we describe the CAPTAIN detector in detail.  Subsequent chapters are dedicated to a more detailed description of the CAPTAIN physics program.

\chapter{The CAPTAIN Detector}

\indent

\begin{figure}[!htb]
\begin{center}
\includegraphics[width=4in,angle=0]{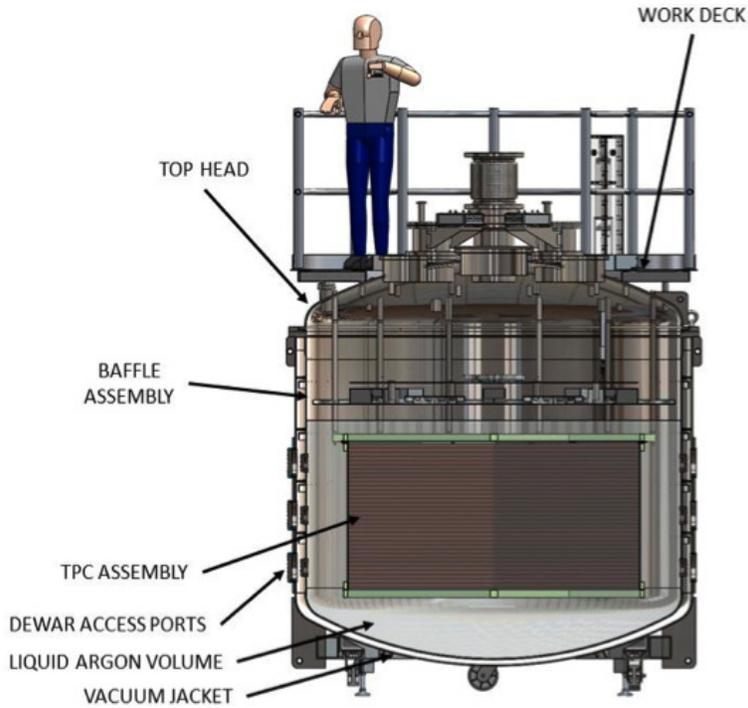}
\caption{\label{CAPTAIN} Schematic drawing of the CAPTAIN detector.} 
\end{center}
\end{figure}

The CAPTAIN program employs two detectors, a prototype detector, mini-CAPTAIN,  and the full CAPTAIN detector.  
The design differences are driven by the cryostat sizes and geometries.  
The prototype is 
smaller and does not have side ports for windows.  
The CAPTAIN detector (Fig.~\ref{CAPTAIN}) is a liquid argon time-projection chamber (TPC) deployed in a portable and evacuable 
cryostat that can hold a total of 7700 liters of liquid argon.  Five tons of liquid argon are instrumented 
with a 2,000 channel TPC and a photon detection system.  The cryostat has ports that can hold optical windows 
for laser calibration and windows for the introduction of charged particle beams.  

The hardware subsystems consist of Cryostats, 
Cryogenics, Elecronics, TPCs, Photon Detection System and 
Laser Calibration System.  
Below, we describe each subsystem in detail.

\section{Cryostats}

The CAPTAIN project utilizes two cryostats for TPC development.  The
first is a small, 1700L vacuum jacketed cryostat provided by UCLA for
the effort.  It is being modified at LANL to provide features to
accommodate and test mini-CAPTAIN.  The vacuum jacket is ~60.25
inches in diameter, and the vessel is ~64.4 inches in height.

The primary CAPTAIN cryostat is a 7700L vacuum insulated liquid argon
cryostat which will house the final TPC.  It is an ASME Section VIII, 
Division 1 U stamped vessel making operation at several national (or 
international) laboratories more straightforward.  The outer shell of the
cryostat is 107.5 inches in diameter, and it is ~115 inches tall.  The
vessel is designed with a thin (3/16 inch) inner vessel to minimize
heat leak to the argon.  All instrumentation and cryogenics are made
through the vessel top head.  The vessel also has side ports allowing
optical access to the liquid argon volume for the laser calibration
system or other instrumentation.  A work deck is to be mounted on the
top head to provide safe worker access to the top ports of the
cryostat.  A baffle assembly will be included in the cold gas above
the liquid argon to mitigate radiation heat transfer from the
un-insulated top head.  Figure \ref{CAPTAIN} shows a schematic of the CAPTAIN
cryostat, TPC, and work deck.

%
%
%

\section{Cryogenics}

Liquid argon (LAr) serves as target and detection medium for the CAPTAIN detector. The argon must 
stay in the form of a stable liquid and must remain minimally contaminated by 
impurities such as oxygen and water.  This is to prevent the loss of 
drifting electrons to these 
electronegative molecules. 
It must also stay sufficiently free of contaminants such as nitrogen to 
avoid absorption of the scintillation light. 

The maximum drift distance is 100.0 cm for the full CAPTAIN detector and 32.0 cm for the prototype. 
To achieve a sufficiently long drift-distance for electrons, 
the O$_2$ contamination is 
required to be smaller than 
750 ppt for mini-CAPTAIN and 240 ppt for CAPTAIN. 
The purity received at 
Los Alamos from industry has an oxygen level of not more than 2.3 ppm. 
Quenching and absorption of 
scintillation light are demonstrated~\cite{ref:WArP, ref:FNAL-N2} to be 
negligible when the N$_2$ 
contamination is smaller than 2 ppm.

\begin{figure}[!htb]
\begin{center}
\includegraphics[width=0.6\textwidth]{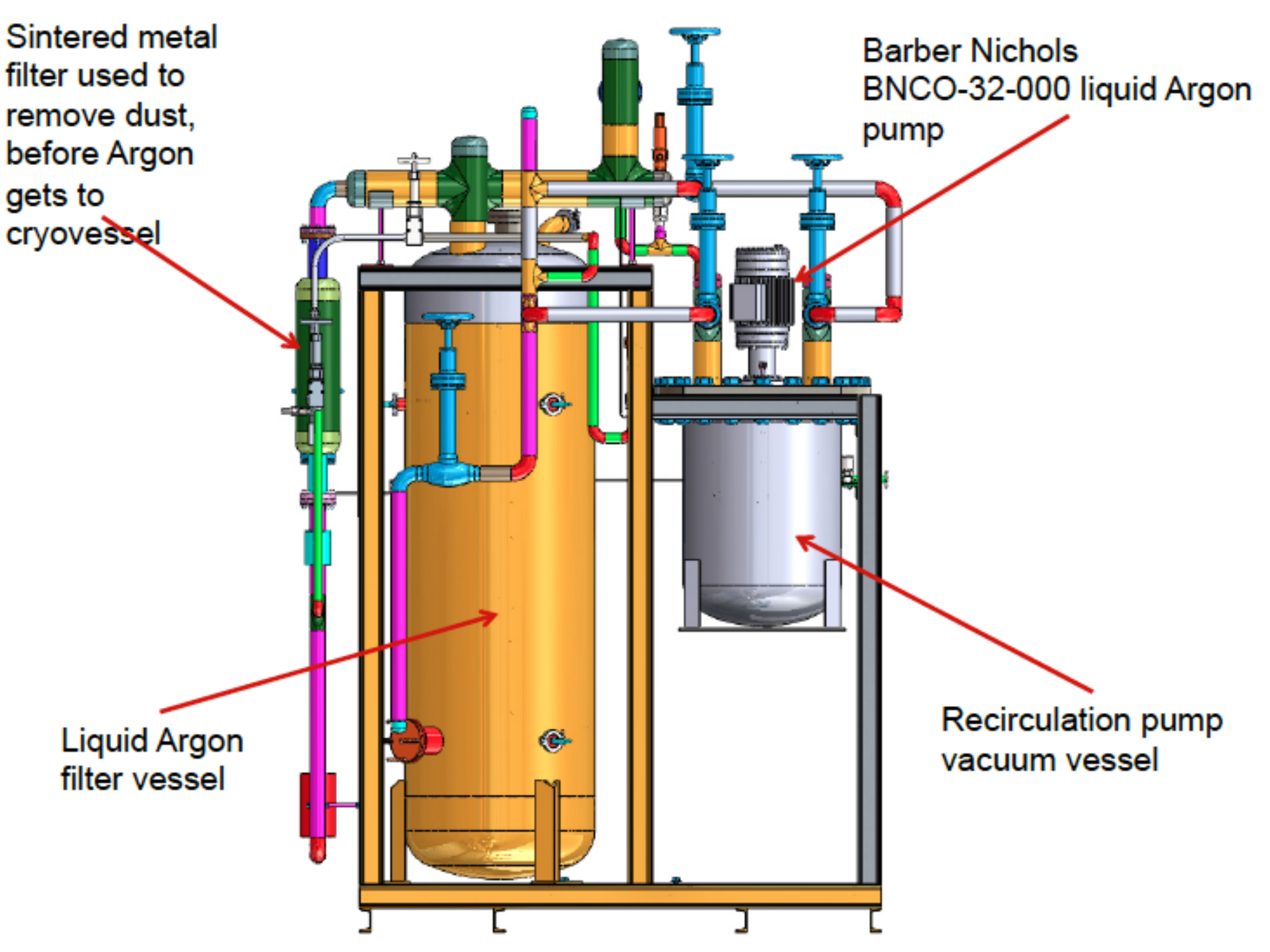}
\end{center}
\caption{The design of purification system~\cite{ref:microboone}~\cite{ref:lapd}.}
\label{fig:cryo}
\end{figure}

The cryogenics system must receive liquid argon from a 
commercial vendor, test its purity, and further purify it. Figure~\ref{fig:cryo} shows the basic design. 
Cryogenic pumps and filter vessels purify the liquid in the detector by removing 
electronegative contaminants. Cryogenic controls monitor and regulate the state of the argon in the 
detector. 
Commercial analytic instruments are used to characterize the oxygen and water 
contaminant levels in the argon. 
The CAPTAIN liquid argon delivery and purification design is 
based on experiences of the 
MicroBooNE experiment~\cite{ref:microboone} 
and the 
Liquid Argon Purity Demonstration (LAPD)~\cite{ref:lapd}, 
both based at Fermilab. 

The CAPTAIN TPC has a liquid argon volume of 7.5 m$^3$, this is equivalent to 6300 m$^3$ 
of argon gas at STP. Assuming a bulk liquid argon contamination level of O$_2$ with a 
1.0 ppm, this is equivalent to 8.253 grams of O$_2$ in the total volume of the 
detector.  
The current design for the vessel that will 
hold the two filter mediums has a total volume of $\sim$80.0 liters, or $\sim$40.0 liters 
for each filter material. 
The dual filter system consists of a bed of molecular sieve 
(208604-5KG Type 4A) to remove moisture and another bed of activated copper 
material (CU-0226 S 14 X 28) to remove oxygen.   
Experience from LAPD shows this should be sufficient.
Both of these filter materials can be reactivated, after reaching 
saturation, by flowing a mixture of argon gas with 2.5\% hydrogen gas at an elevated temperature. 

The design 
utilizes a 10-12 gal/min capacity commercial centrifugal pump.  
Magnetic coupling prevents contamination through shaft seals. 
The pump will be mounted with the filter vessel on a single skid in 
order to achieve portability.  

A sintered metal filter is used to remove dust from the 
liquid argon prior to its delivery to the cryovessel.

\section{Electronics}

\begin{figure}[!htb]
\begin{center}
\includegraphics[width=0.6\textwidth]{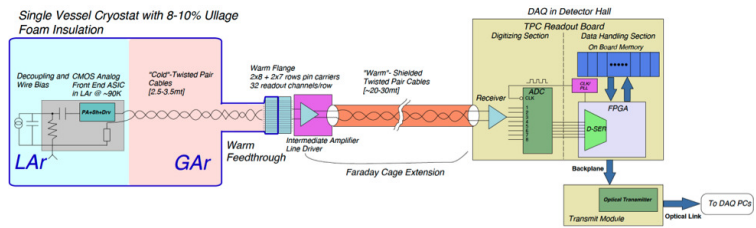}
\end{center}
\caption{The MicroBooNE electronics chain to be used for CAPTAIN~\cite{ref:microboone}.}
\label{fig:electronics}
\end{figure}

The electronic components for the TPC are identical to those of the
MicroBooNE experiment at FNAL~\cite{ref:microboone}.  A block diagram of the
electronics is in Figure \ref{fig:electronics}.  The front-end mother board is designed
with twelve custom CMOS Application Specific Integrated Circuits
(ASIC).  Each ASIC reads-out 16 channels from the TPC.  The mother
board is mounted directly on the TPC wire planes and is designed to be
operated in liquid argon.  The output signals from the mother board
are transmitted through the cold cables to the cryostat feed-thru to
the intermediate amplifier board.  The intermediate amplifier is
designed to drive the differential signals through long cable lengths
to the 64 channel receiver ADC board.  The digital signal is then
processed in an FPGA on the Front End Module (FEM) board.  All signals
are transmitted via fiber optic from a transmit module to the data
acquisition computer.

\section{TPC}

\subsection{CAPTAIN TPC}

The TPC consists of a field cage in a hexagonal shape with a mesh
cathode plane on the bottom of the hexagon and a series of four wire
planes on the top with a mesh ground plane.  
The apothem of the TPC 
is 100 cm and the drift length between the anode
and cathode is 100 cm.  
In the direction of the electron drift, there are four wire planes.  
In order, they are the a grid, U, V, and collection (anode)
plane.  
The construction material of the TPC is FR4 glass fiber
composite.  
All wire planes have 75 $\mu$m diameter copper beryllium wire
spaced 3 mm apart and the plane separation is 3.125 mm.  Each wire
plane has 667 wires.  
The U and V planes detect the induced signal
when the electron passes through the wires.  
The U and V wires are
oriented $\pm$60 degrees with respect to the anode wires.  
The anode wires
measure the coordinate in direction of the track and U and V are
orthogonal to the track.  The third coordinate is determined by the
drift time to the anode plane.

\begin{figure}[!htb]
\begin{center}
\includegraphics[width=0.6\textwidth]{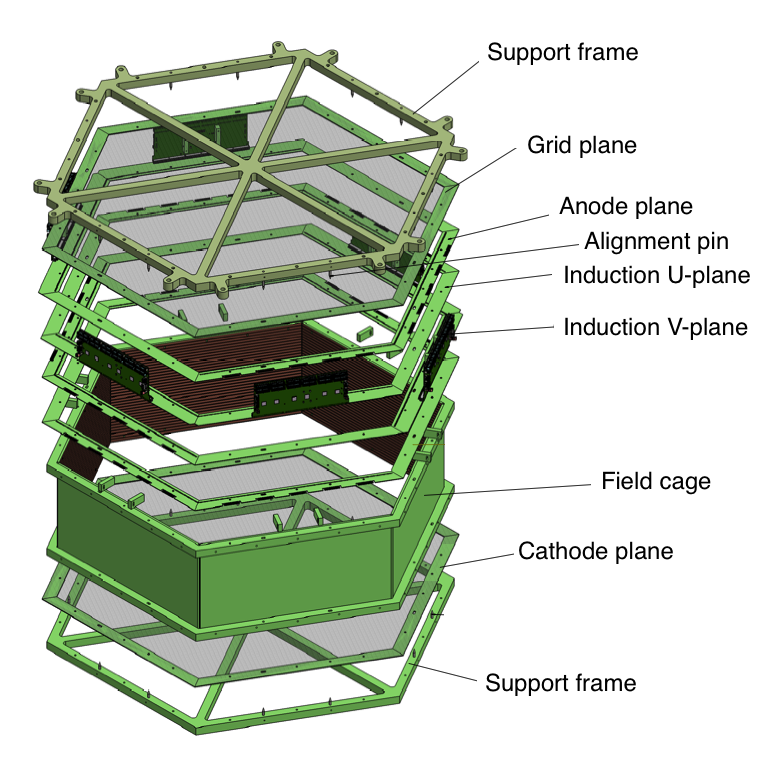}
\end{center}
\caption{Component detailed view of the CAPTAIN TPC.}
\label{fig:tpc}
\end{figure}

The field cage is realized in two modules: the
drift cage module, and the wire plane module.  The wire plane module
incorporates a 2.54 cm thick FR4 structural component that supports
the load of the four wire planes so that the wire tension is
maintained.  The field cage is double sided gold plated copper clad
FR4 arranged with 5 mm wide traces separated by 1 cm.  A resistive
divider chain provides the voltage for each trace.  The design voltage
gradient on the divider chain is 500 V/cm when 50 kV is applied to the
cathode.  
The electrons from the ionized event are collected on the
anode plane.
The U, or V planes detect signals via induction, and are made 
transparent to electrons via biasing.
The drift velocity of the 
electrons with 500 V/cm is 1.6 $mm/\mu s$.  See Figure~\ref{fig:tpc}

\subsection{Prototype TPC}

The prototype TPC is a smaller version the CAPTAIN TPC.  The drift
length is 32 cm and the apothem is 50 cm.  Each
wire plane has 337 wires.  The prototype is designed to test the
mechanical construction details of the TPC, the cold electronics, and
the back-end data acquisition system prior to the construction of the
full scale CAPTAIN.  It also allows the early development of the data
acquisition software so that CAPTAIN can produce data as soon as the
hardware is ready.  It will also provide the needed operational
experience to run the full scale CAPTAIN.

\section{Photon Detection System}

By detecting the scintillation light produced during interactions in the CAPTAIN detector, the photon detection system provides valuable information.  Simulations show that detection of several photoelectrons per MeV for a minimum ionizing particle (MIP) in a TPC with a field of 500 V/cm improves the projected energy resolution of the detector by 10-20\%.  
Such improvement stems from the anti-correlation between the production of scintillation 
photons and ionization electrons, a phenomenon which has been conclusively observed 
to improve calorimetry already in liquid xenon~\cite{ref:conti}.  
Hints of it have already been seen 
in our own re-analysis of older liquid argon data that included simultaneous measurements 
of light and charge yields at the same electric fields~\cite{ref:doke}.   
If confirmed by CAPTAIN, it will increase the utility of the photon detection 
systems of other experiments such as LBNE, as well as argon-based dark matter 
detectors.  Just as with the charge signal, the amount of light produced by a 
particle traversing argon is a function of the energy deposited.  
The scintillation light can also be used to determine the energy of neutrons from time of flight when the experiment is placed in a neutron beamline 
by giving the time of the interaction with few nanosecond resolution. 

Liquid argon scintillates at a wavelength of 128 nm which unfortunately is readily absorbed by most photodetector window materials.  It is thus necessary to shift the light to the visible.  The photon detection system is composed of a wavelength shifter covering a large area of the detector and a number of photodetectors to collect the visible light.  The baseline CAPTAIN photon detection system uses tetraphenyl butadiene (TPB) as a wavelength shifter and sixteen Hamamatsu R8520-500 photomultiplier tubes (PMT) for light detection.  The R8520 is a compact PMT approximately 1'' x 1'' x 1'' in size with a borosilicate glass window and a special bialkali photocathode capable of operation at liquid argon temperatures (87 K).  It has a 25\% quantum efficiency at 340 nm.  TPB is the most commonly used wavelength shifter for liquid argon detectors and has a conversion efficiency of about 120\% when evaporated in a thin film.  It has a re-emission spectrum that peaks at about 420 nm \cite{bibTPB}.  The TPB will be coated on a thin piece of acrylic in front of the PMTs.    Eight PMTs will be located on top of the TPC volume and eight on the bottom.  This will provide a minimum detection of 2.2 photoelectrons per MeV for a MIP.  The amount detected will increase if the entire top and bottom surfaces are coated with TPB.  

The PMTs will use a base with cryogenically compatible discrete components.  The cable from the base to the cryostat feedthrough is Gore CXN 3598 with a 0.045" diameter to reduce the overall heat load.  The PMT signals will be digitized at 250 MHz using two 8-channel CAEN V1720 boards.  
The digitizers are readout through fiber optic cables by a data acquisition system written for the MiniCLEAN experiment \cite{bibGastler}.

The CAPTAIN detector will serve as a test platform for the evaluation of 
alternative photon detection system designs.  
The design will allow testing options in a operating TPC 
with cosmic muons and in various beamlines.  
Such options include other wavelength shifting films, 
acrylic light guides or 
doped panels to the photodetectors, and other types of 
photodetectors such as SiPMTs, larger cryogenic PMTs, 
and avalanche photodiode arrays.  
In addition, CAPTAIN can test methods of calibrating the 
photon detection system with the laser calibration system 
or alternatively a series of UV and blue LEDs.

\section{Laser Calibration System}

The first measurement of photoionization of liquid Argon was performed by Sun et al.~\cite{ref:SunLaser}. 
Using frequency quadrupled Nd-YAG laser to generate 266nm light the authors demonstrated that the ionization 
was proportional to the square of the laser intensity.  The ionization potential of liquid Argon is 13.78 eV, 
slightly lower than the energy of 3 photons from a 266nm quadrupled Nd-YAG laser.  The ability to create 
well-defined ionization tracks within a liquid Argon TPC provides an excellent calibration source that can 
be used to measure the electron lifetime in-situ and to determine the drift field within the TPC itself.  
Significant progress has been made in this field and is documented in Rossi et al.~\cite{ref:RossiLaser}.   
The CAPTAIN TPC provides an excellent test bed for a future LBNE laser calibration system.

\begin{tabular}{|c|c|c|c|}
\hline
Wavelength & 1064 nm & 532 & 266 \\
\hline
Pulse Energy &  850 mJ   &  400 mJ & 90 mJ \\
Pulse Duration & 6 ns & 4.3 ns & 3 ns \\
Peak Power & 133 MW & 87 MW & 28 MW \\
Peak Intensity & 1500 $GW/cm^2$ & 985 $GW/cm^2$ & 317 $GW/cm^2$ \\
Photon Energy & 1.17 eV & 2.33 eV & 4.66 eV \\
Photon Flux & $ 8 x 10 ^{30} \gamma/(s-cm^2)$ & $ 2.6 x 10 ^{30} \gamma/(s-cm^2)$ & $ 0.42 x 10 ^{30} \gamma/(s-cm^2)$ \\
\hline
\end{tabular}

To avoid surface irregularities that may disperse the laser beam the CAPTAIN TPC will employ optical 
access on the sides of the detector (Fig.~\ref{CAPTAIN}).  A LANL existing 
Quantel �Brilliant B� Nd-YAG laser will be used to 
ionize the liquid Argon.  The laser parameters are given in Table 1.  The laser and mirrors are in hand 
and the design of the mirror mounting system on the TPC frame has begun.  The design seeks to be 
flexible and allow several paths through the liquid Argon, including parallel and at an angle to the wire 
plane.  This will allow us to determine the electron lifetime within the CAPTAIN TPC.  

\section{Special Run Modes}

\subsection{Tests of Doping Liquid Argon to Improve Light Output}

Previous research \cite{bibKubota1, bibKubota2, bibPollman} suggests
that there is a potential benefit to doping liquid argon with xenon or
other wavelength shifting compounds to improve the collection of
photons in a LAr TPC with little effect on the ionization readout.
The possible advantages would be to shorten the triplet state lifetime
for the scintillation photons from 1.6 microseconds and possibly shift
the scintillation light from 128 nm to 178 nm (for xenon) or higher
(for other compounds).  A shift in scintillation light wavelength
would have a large impact since the Rayleigh scattering length is
proportional to ${1/\lambda^4}$ resulting in less scattering and better
  time resolution in a large detector.  Higher wavelength
  scintillation light would also open up the possible use of other
  photodetectors and remove the need for wavelength shifting coatings
  such as TPB.  CAPTAIN would serve as an ideal detector to study how
  much xenon or dopant would be needed to speed up the triplet
  lifetime.  The literature has studied a broad range of levels from
  several ppm to $\sim$1\% but only in small detectors with a poor
  ability to determine the final mixture.  How the xenon or dopant
  remains in the LAr over time could be examined along with the
  ability to achieve uniform mixing in a large detector.  
  With the TPC, the affect of concentration on the drift
  velocity and electrostatic properties would be examined.  Finally,
  the best method for introducing the xenon or dopant could be developed.

%
%

\chapter{Neutrons}

\indent

\section{Physics Importance}

A detailed understanding of neutron interactions with argon is crucial for the success of LBNE.  They impact two major LBNE missions:  low-energy neutrino detection -  important for supernova neutrino studies, and neutrino oscillation studies with medium-energy neutrinos.  

In the first case, neutron spallation on argon nuclei is an important channel for the production of isotopes that comprise the background to the detection of low-energy neutrinos  - for example, those from supernova bursts.  
Studying neutron spallation of the argon nucleus with a well-characterized neutron beam is therefore compelling.  Additionally, 
the neutral-current interactions of supernovae neutrinos on argon nuclei will leave them in excited states. 
This interaction can be well-simulated by bombarding argon with 
fast neutrons. 
The detection and identification of de-excitation events following neutron-argon interactions is an important step in establishing whether or not neutral-current interactions of supernovae neutrinos are detectable in a LAr TPC.  

\begin{figure}
\begin{center}
\includegraphics[scale=0.4,angle=0]{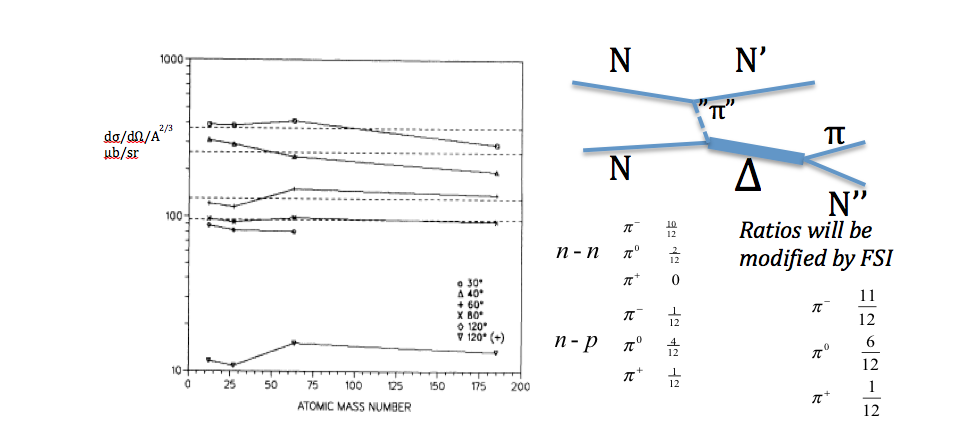}
\caption{\label{PionRatio} number of pions produced by n-n and n-p interactions and their ratios from~\cite{ref:Brooks}.} 
\end{center}
\end{figure}

In the second case, neutrons are an important component of the hadronic system in medium-energy neutrino interactions with argon.  
Charged hadrons are well-measured and neutral pions will decay quickly into 
gamma-rays that point back to the neutrino vertex.  Neutrons, on the other hand, 
will travel some distance from the neutrino vertex before interacting and will 
complicate the reconstruction of these events. 
Besides this, the study of neutron-induced pion production and spallation events in 
LAr in terms of their topology, of the multiplicity and identity of the visible particles in 
the final state and of their kinematic properties, is important for LBNE because similar 
events will be produced by neutrino and antineutrino interactions.
Additionally, 
at the near site, neutrons will comprise an important in-time background to 
neutrino detection.  Measuring neutron interactions as a function of neutron 
energy up to relatively high kinetic energies is thus important.  With these 
topics in mind, we have developed a neutron running program with CAPTAIN 
in addition to measurements at neutrino beams. Such measurements have not been previously performed - they are unique for this program.

For few GeV neutrinos (antineutrinos) delta production is the principle source of pion production. 
In neutral current interactions only $\Delta^+$ and $\Delta^0$ will be produced. 
However the neutrino cross section is larger by a factor of $\sim 2$ because of differing interference effects due to the opposite helicity of neutrinos and antineutrinos. 
The final charge state of the pions produced in neutral current interactions is important for identifying the relative neutrino and antineutrino flux. 
Pions are readily absorbed and can change their charge state via final state interactions, so it is also important to characterize the final state interactions in argon.  
Figure \ref{PionRatio} left (reproduced from~\cite{ref:Brooks}) shows the cross-section 
for pion production from 450 MeV neutrons at various angles from different nuclei scaled by $A^{2/3}$. 
The upper 5 distributions are for $\pi^-$ while the one at the bottom is for $\pi^+$. 
The large dominance of $\pi^-$ over $\pi^+$ might be surprising but is qualitatively understandable if one considers pion production as proceeding via delta production as shown in Figure \ref{PionRatio} right. 
The figure on the right illustrates the asymmetry in pion production by neutrons on a N=Z nucleus via the delta resonance. 
In reality, the observed ratio $\pi^-/\pi^+$ is always less than 11 indicating the 
importance of the role played by final state interactions. 
It will be important to separate the pions coming from deltas formed in the nucleus from those formed on the incident neutron as the former better reflect the deltas formed via neutrinos (antineutrinos).

\begin{figure}
\begin{center}
\includegraphics[scale=0.4,angle=0]{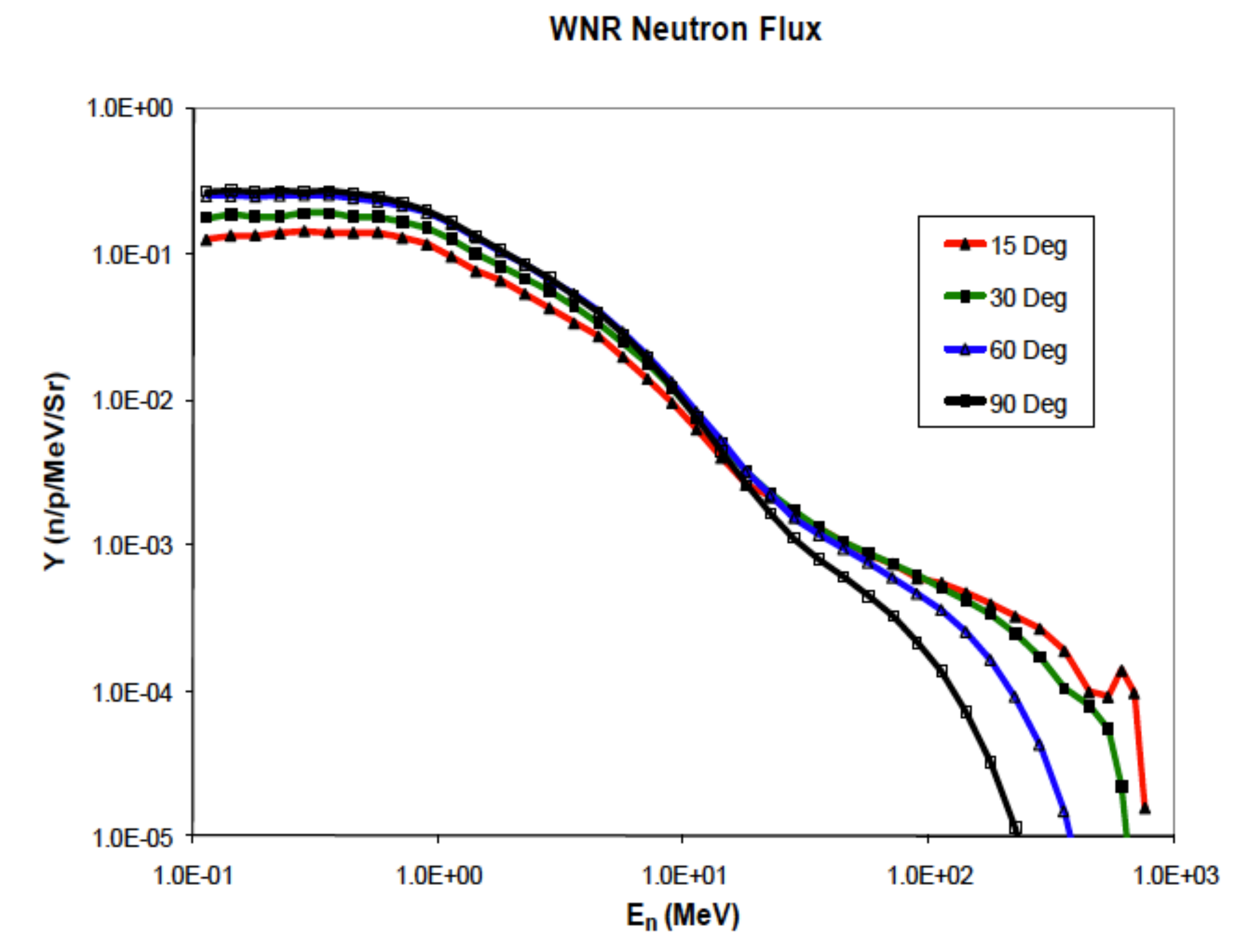}
\caption{\label{NeutronFluxWNR} The neutron flux at the LANSCE WNR facility.  It is anticipated CAPTAIN would run in the 15R beamline (i.e. 15 degrees off of center).} 
\end{center}
\end{figure}

The CAPTAIN program takes advantage of the proximity of the Los
Alamos Neutron Science Center (LANSCE) to the CAPTAIN commissioning hall.  LANSCE 
has a beamline with a well-characterized
neutron energy spectrum with an endpoint close to 800 MeV kinetic energy (Figure~\ref{NeutronFluxWNR}). 
The energy of the incoming neutrons can be determined on an event-by-event basis by measuring their time of flight.

It is worth noting, while the goal for LBNE is to install the far detector deep underground, 
the current approved scope for LBNE has the far detector on the surface.  
For surface operations, 
collection and analysis of neutron data are absolutely critical.  
First, at the surface, there is a significant cosmic-ray 
neutron flux that will impinge on the detector.  The spectrum shown in Figure~\ref{NeutronFluxWNR} 
is quite similar to the cosmic-ray spectrum, but much more intense.  With a few days of running, we will measure the neutron production of isotopes such as $^{40}Cl$ that constitute an important background to supernova neutrino detection.  Surface running also presents a challenge to the long-baseline neutrino program.  High-energy cosmogenically produced neutrons can produce events 
with neutral pions that can mimic the electron neutrino appearance signal for LBNE.  
Currently, simulations show that ~10\% of the electron neutrino appearance background could come from fast neutrons with complicated FSI, but uncertainties are large and must be measured prior to finalizing surface shielding requirements and photon system specifications.

In the following we briefly describe the measurements that we plan to perform using CAPTAIN 
in the LANSCE neutron beam.

\section{High-intensity neutron running}

The neutron flux shown in Figure~\ref{NeutronFluxWNR} is much more intense than that 
produced by cosmic-ray interactions at the LBNE far detector site, so a single day of running 
will produce years worth of neutron spallation events in CAPTAIN.  
We will run in an integrated mode where we expose the detector to the full intensity of the 
beam, close the shutter, and observe the decay of isotopes such as $ ^{40}Cl$.  
We are currently investigating making measurements in neutron beamlines with dedicated 
detectors such as GEANIE (GErmanium Array for Neutron Induced Excitations) 
for high-precision measurements of production cross-sections that will be input to 
simulations in LBNE and will cross-check the measurements made in CAPTAIN to 
determine the efficiency.

\section{Low-intensity neutron running}

Low-intensity running allows us to correlate specific topologies with neutron kinetic energy via 
time of flight.  
Although they are named ''low-energy neutron run'' and ''high-energy neutron run'' both 
measurements will be performed at the same time, given the wide range of the continuous 
energy spectrum of the incoming neutrons.

\subsection{Low-energy neutron run}
The goal of a low intensity, low-energy neutron run is to measure an excitation spectrum in $^{40}Ar$ and to study the reconstruction capabilities of $^{40}Ar^*$ de-excitation events in a liquid argon TPC.

The detection of a galactic supernova neutrino burst in LBNE requires the capability to tag and identify the following charged-current (CC) and neutral-current (NC) interactions:
\begin{equation}
\begin{split}
\nu_e+^{40}Ar\rightarrow e^- +^{40}K^*  (CC)\\
\nu_x+^{40}Ar\rightarrow \nu_x +^{40}Ar^* (NC)
\end{split}
\end{equation}
In order to gain insight into the neutral-current interaction we propose to place the CAPTAIN detector in the neutron beam at LANSCE and measure:
\begin{equation}
n+^{40}Ar\rightarrow n +^{40}Ar^*
\end{equation}

This interaction will provide a test bed for identifying $^{40}Ar^*$ de-excitation events inside a liquid Argon TPC.  The relationship between neutron-induced and neutrino-induced interactions on $^{40}Ar$ will be investigated by comparing the neutron beam data to Monte Carlo simulations of both types of events.

\subsection{High-energy neutron run}
The goal of a low-intensity, high-energy neutron run is to study neutron induced events in order to characterize their topology, multiplicity, and identity of the visible particles produced along with their kinematic properties.  We plan to compare our results with existing models, improve them, and use them to simulate LBNE events. 

The experiment will use neutrons above $\sim 400$ MeV, where pion production can occur and the relevant events can be clearly seen in a LAr TPC.  The differential cross section of pion production on C, Al, Cu, and W has been measured . The observed $A^{2/3}$ dependence of the cross sections shows that these cross sections can be readily predicted in argon, even if final states are very uncertain. 
In LAr the $\pi^-$ will be absorbed on an argon nucleus leading to a variety of multi-nucleon final states.

Brooks et al.\ \cite{ref:Brooks} did not observe anything beyond the pion, so measurements with CAPTAIN will provide further information on details of the interaction.  
CAPTAIN will also measure spallation events in the neutron beam and try to measure the effect of these events on LBNE electron neutrino appearance backgrounds.

\section{Run Plans}

We anticipate neutron running in early FY15.  The 2015 run cycle begins in August of 2014 and continues 
through early calendar 2015.  Proposals are due to the LANSCE PAC in mid-April of 2014, so we will 
prepare our proposals during FY14.

\chapter{Neutrinos}

\indent

Beyond neutron running, there are several possibilities for neutrino running beginning 
in FY15 or FY16.
Two promising possibilities are running in an on-axis position in the NuMI beamline
and running with the neutrino flux created at the Spallation Neutron Source.  
We have carefully designed the cryostat and cryogenics systems such that they can 
be transported and operated in a variety of facilities with a minimum of safety or operational 
challenges.  For example, the cryostat is ASME U-stamped.

\section{Running at NuMI}
 
LBNE will measure neutrino oscillation phenomena with a baseline of 1300 km 
using, primarily, the first oscillation maximum.  At that baseline, the neutrino 
energies in the first maximum range from ~ 1.5 to 5 GeV.  
Neutrino cross-sections are poorly understood on any nuclear target in this 
energy regime.  For argon, the ArgoNEUT collaboration has produced the first 
and only inclusive cross-section measurement in the energy regime important 
for LBNE with 379 events~\cite{ref:argoneut} integrated over the neutrino 
spectrum produced by the NuMI low-energy tune.
LBNE must use the {\it full} CC cross-section for the oscillation analysis and thus 
must have robust methods to determine the neutrino energy.  
A detailed study of interactions in the energy regime corresponding to the LBNE 
first maximum is crucial.  {\it The experiment simply will not work without it.}

\begin{figure}
\begin{center}
\includegraphics[scale=0.5, angle=0]{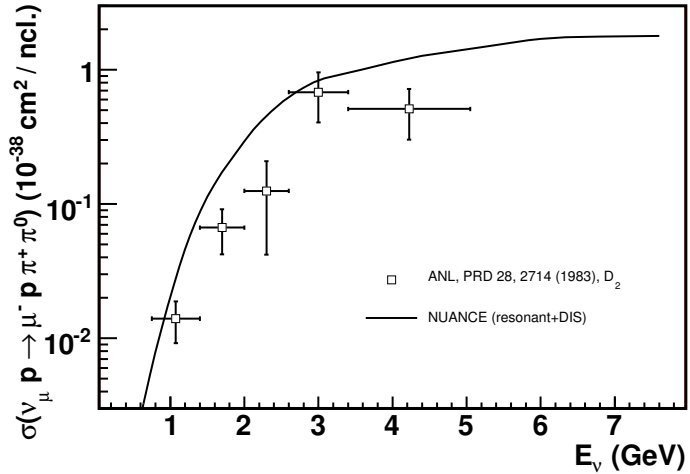}
\caption{\label{fig:zellerformaggio} Existing measurements of the 
$\nu_{\mu} p \rightarrow \mu^{-}  p  \pi^{+} \pi^{0}    $ cross-section as a function of 
neutrino energy reproduced from Figure 24 of Formaggio and Zeller~\cite{ref:forzel}.    }
\end{center}
\end{figure}

Figure~\ref{fig:zellerformaggio} shows the state of the art for the exclusive channel: 
$\nu_{\mu} p \rightarrow \mu^{-}  p  \pi^{+} \pi^{0}    $ on a free nucleon~\cite{ref:forzel}.  
Clearly more data are crucial in an era of precision neutrino physics.  

\subsection{On-axis running in NuMI}
 The NuMI beamline was constructed for the MINOS experiment and 
 will be running with the medium energy tune to support the No$\nu$a 
 and Miner$\nu$a experiments.  The medium energy tune will provide 
 an intense neutrino beam with a broad peak between approximately 
 1 and 10 GeV.  

The NuMI running of the CAPTAIN detector in both a neutrino and
antineutrino beam is an integral part of understanding the neutrino
cross sections needed by LBNE, and liquid argon detectors in general,
to interpret neutrino oscillation signals. Measurements of CAPTAIN in an
on-axis position in the NuMI beam are complementary to low-energy neutrino 
measurements
made using MicroBooNE; moreover, with a fiducial mass approximately 20
times larger than the ArgoNEUT detector, CAPTAIN will contain the hadronic
system for a significant fraction of events.  CAPTAIN will make high
statistics measurements of neutrino interactions and cross-sections in
a broad neutrino energy range, from pion production threshold to deep
inelastic scattering.
 
The fine resolution of the detector will allow detailed studies of low
energy protons that are often invisible in other neutrino detector
technologies.  In addition, liquid argon TPC's give good separation
between pions, protons and muons over a broad momentum range.  With
these characteristics, CAPTAIN will make detailed studies of
charged-current (CC) and neutral-current (NC) inclusive and exclusive
channels in the important and poorly understood neutrino energy range
where baryon resonances dominate.  The first oscillation maximum 
energy at LBNE (2-5 GeV) 
is similarly dominated by baryon resonances.  
The exclusive channels CAPTAIN will measure include single charged and
neutral pion production and single photon production.  Measurements
near strange production threshold will also be made.
 
\begin{figure}
\begin{center}
\includegraphics[scale=0.4, angle=0]{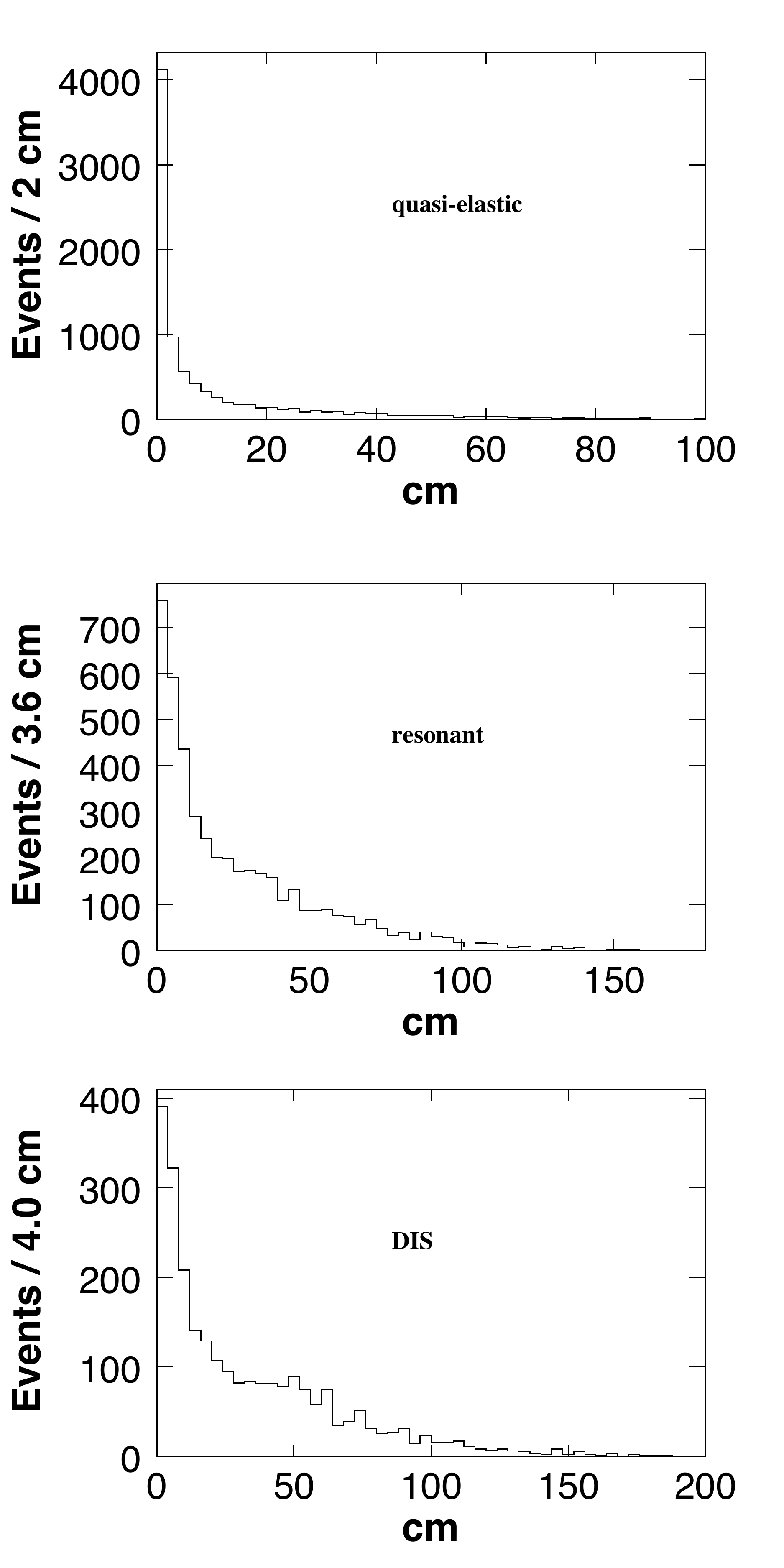}
\caption{\label{CAPTAINandNUMI} Each graphic shows the furthest distance any charged particles or electromagnetic 
showers travelled in the detector from the selection of contained events.  The upper left figure shows quasielastic events.  The upper right figure shows the same from events created via nucleon resonances.  The lower figure shows the same from events created by deep inelastic scattering.  The detector contains about 10 \% of all interactions where ``contained'' is as defined in the text.}
\end{center}
\end{figure}

Since the NuMI running comes after the neutron running at LANSCE,
we will employ the neutron-interaction identification techniques developed
in CAPTAIN in the reconstruction of the energy of the hadronic system.
 
In addition to the physics program for this detector, running in a neutrino beam with
similar characteristics to the proposed LBNE beam will provide validation of its
technology, including exclusive particle reconstruction and identification, shower
reconstruction, and reconstruction of higher energy neutrino events with significant particle multiplicities.
 
Finally, in order to accommodate neutrino running, a simulation of neutrino interactions with argon with a neutrino
energy spectrum of the on-axis, medium-energy NUMI tune was carried out.
It was determined that the current geometry
would contain 10\% of all neutrino events where containment is defined to be ``all but lepton and neutrons.''
Depending on achieved POT, this will yield roughly 400,000 contained events per year. Distributions of the particle
that travels the furthest from the vertex (with the exception of neutrons and leptons) is shown in Figure~\ref{CAPTAINandNUMI}.

\subsection{Off-axis running in NuMI}
We may be able to deploy the CAPTAIN detector in front of the No$\nu$a near detector in an off-axis 
position.  The off-axis NuMI beam provides a relatively narrow neutrino energy peak at about 2 GeV.  
Such running would provide a wealth of information at a specific neutrino energy close to the LBNE 
oscillation maximum.  While interesting in its own right, the information would serve as a valuable 
lever arm for interpreting the broad-band on-axis data.  Studies of this option are just beginning.

\section{Running at the SNS}

The measurement of the time evolution of the energy and flavor spectrum of neutrinos from supernovae can 
revolutionize our understanding of neutrino properties, supernova physics, and discover or tightly constrain 
non-standard neutrino interactions.  LBNE has the capability to make precise measurements of supernova 
neutrinos.   For example, collective neutrino oscillations imprint distinctive signatures on the time evolution 
of the neutrino spectrum that depend, in a dramatic fashion, on the neutrino mass hierarchy and mixing 
angle $\theta_{13}$.  Current knowledge of the neutrino argon cross sections and interaction products at 
the relevant energies ($<$50 MeV) (there are no neutrino measurements in this energy range) limit the 
ability of detectors to extract information on neutrino properties from a supernova neutrino burst.  Cascades 
of characteristic de-excitation gamma rays are expected to be associated with different interaction channels, 
which could enable flavor tagging and background reduction.  Currently the ability of LArTPC detectors to 
observe these gamma rays (and accurately reconstruct their energy) is very uncertain.  
CAPTAIN will afford a nearly unique opportunity 
to measure key neutrino-nuclear cross sections in both 
the charged and neutral current channels that would allow us to make better use of the supernova neutrino 
burst signal.

The observation of 11 (mostly) anti-neutrino events from SN1987A confirmed the general model of supernovae explosions, demonstrating that the bulk of the gravitational binding 
energy, $ 8 \times 10^{52} $ ergs is released in the form of neutrinos.  The LBNE detector with 
34 kT of fiducial mass and an Argon target would detect more than 1000 events from a supernova at 10 kpc.  
There are four processes that can be used to detect of supernova neutrinos in a liquid Argon detector:
\begin{equation}
\nu_e +   ^{40}Ar   \rightarrow   e^{-}  +  ^{40}K^{*}
\end{equation}
\begin{equation}
\bar{\nu_{e}}  +   ^{40}Ar   \rightarrow   e^{+}  +  ^{40}Cl^{*}
\end{equation}
\begin{equation}
\nu_x +   e^{-}   \rightarrow   \nu_x  +  e^{-} 
\end{equation}
\begin{equation}
\nu_x +   ^{40}Ar   \rightarrow        ^{40}Ar^{*}  + \gamma
\end{equation}

The vast majority of these neutrinos would be detected via the first process above.  
Though small in number, the elastic scattering reaction preserves the neutrino direction, enabling localization of the direction of the supernova and the neutral current reaction 
allows for a calibration of the total energy released in neutrinos.  
While the elastic scattering cross section has been measured, the 
charged current reactions in argon have only theoretical predictions.  
In addition, while elastic scattering can also be measured in water 
Cherenkov detectors, the argon CC interaction is unique in 
that it has a large cross-section and has the potential to provide a 
better handle on the energy of the initial neutrino.
The estimated theoretical uncertainty in the cross section calculations 
is stated to be 7\%, however the authors note that a small change in the 
Q value of the excited argon state could lead to a 10-15\% change in the cross section.

\begin{figure}[!htb]
\begin{center}
\includegraphics[scale=0.75,angle=0]{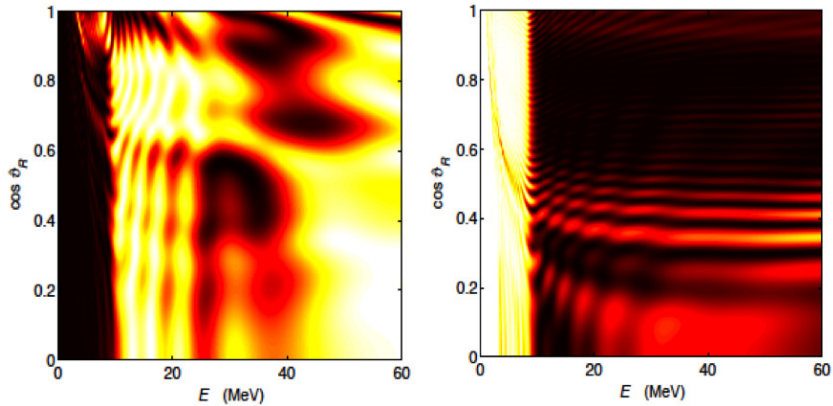}
\caption{\label{SNswap} Spectral swap of neutrino energy spectrum between normal hierarchy (left panel) and inverted hierarchy (right panel).  The y-axis is the emission angle of the neutrino, which is not observable.  The observed energy spectrum is the projection of these figures onto the x-axis.  Figure from reference \cite{ref:duan2010} .} 
\end{center}
\end{figure}

It has recently been realized that the evolution of the neutrinos as they leave the 
protoneutron star surface is more complicated than previously believed.  Collective 
oscillations of the late time ( approximately 10-20 seconds after the neutronization burst) 
neutrinos leads to a spectral swap~\cite{ref:duan2010} 
shown in Figure \ref{SNswap}.  The figure shows the probability that a neutrino of species x would survive without oscillating to another species as it propagates through the neutrinosphere.  This survival probability is shown as a function of the neutrino energy (x-axis) and the emission angle from the protoneutron star (y-axis).  The left panel is for the normal mass hierarchy and the right panel is for an inverted neutrino mass hierarchy.  What can be seen is that in the normal hierarchy, all neutrinos with energy less than 10 MeV oscillate to a different species and all those above 10 MeV survive as the same species.  For an inverted hierarchy it is just the opposite (the low-energy neutrinos survive without oscillation).  Since the temperature (or energy spectrum) of the neutrinos is dependent upon the neutrino flavor, this spectral swap could be observed in a detector that is sensitive to electron neutrinos.

Extracting the physics from the detection of a Galactic supernova depends upon understanding of the neutrino argon cross sections, the ability to distinguish the neutrino charged current reaction given above from the anti-neutrino charged current reaction (which leads to an excited Cl nucleus), the ability to isolate the neutral current reaction, and the energy resolution of the detector.  

To extract the neutrino physics from the detection of a supernova burst one needs to convert the measured electron neutrino spectrum to a source flux (as a function of energy) and to measure the complete (over all neutrino species) neutrino energy distribution.  This will require accurate knowledge of the charged current cross section for converting Ar to K and the neutral current cross section for creating the excited $^{40}Ar$ state.  In addition one needs to clearly tag such events, which can be done by detecting (and measuring the energy of) the de-excitation gamma rays as the excited states of K, Ar  or Cl decay. 

We propose to run the CAPTAIN (Cryogenic Apparatus) LAr TPC at the SNS to:
\begin{itemize}
\item Measure the neutrino argon charged current cross sections in the energy region of interest.   
\item Investigate the capability of a liquid argon TPC to measure the de-excitation gamma rays from the excited nuclear states of Ar, Cl, and K).
\item Measure the energy resolution of a liquid argon TPC at low energies in a realistic environment similar to operation of LBNE at the surface, to demonstrate that one can properly tag the events.
\end{itemize}

\begin{figure}[!htb]
\begin{center}
\includegraphics[scale=0.75,angle=0]{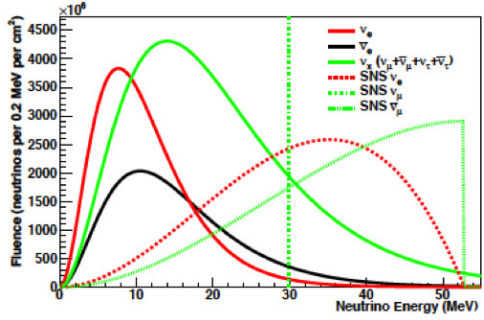}
\caption{\label{nspectrum} Neutrino energy spectra from supernova bursts neutrinos (solid lines) and the neutrino energy spectra from the stopped pion source at the SNS.  Figure from reference \cite{ref:Bolozdynya2013} } 
\end{center}
\end{figure}

The Spallation Neutron Source in Oak Ridge, TN, provides a high-intensity source of neutrinos from stopped pions in a mercury target.  The energy spectrum of the neutrinos from a stopped pion source is well known, 
has an endpoint near 50 MeV, and covers the energy range of supernova burst neutrinos.  
Figure \ref{nspectrum}  shows the SNS neutrino energy spectra and that of supernova burst neutrinos.  
The timing characteristics of this source will help reduce the neutron background, though shielding will 
most likely be required.

\begin{figure}[!htb]
\begin{center}
\includegraphics[scale=0.5,angle=0]{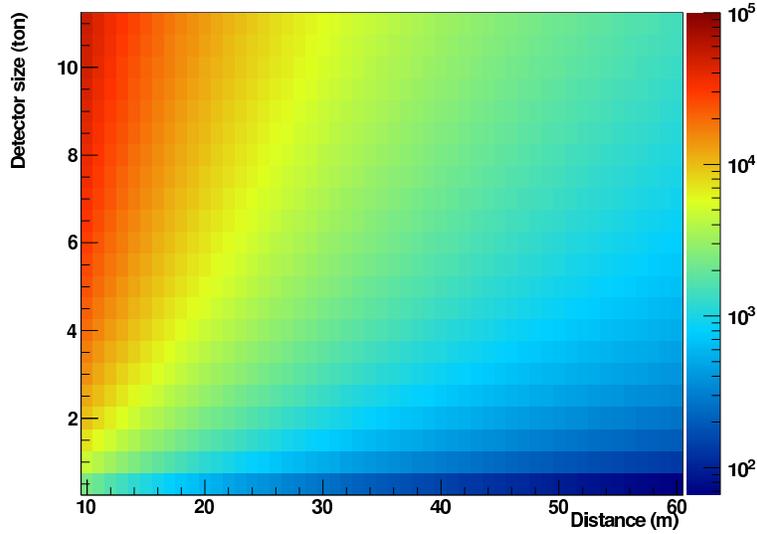}
\caption{\label{SNSEventRates} Estimated event counts per year in argon as a function of detector mass and distance from the SNS target \cite{ref:Bolozdynya2013}.} 
\end{center}
\end{figure}

The interaction rates in argon are shown in Figure \ref{SNSEventRates} as a function of distance and argon mass.  
A five-ton detector would measure thousands of events per year if sited sufficiently close to the target.   Neutrino interaction cross-sections in argon and interaction product distributions, including de-excitation gammas, could be measured for the first time.  Such an experiment would also be valuable for understanding the response of a LArTPC detector to neutrinos in this energy range.

\section{Stopped Pion Source at the BNB}

The Booster Neutrino Beam (BNB) at FNAL was designed and built as a conventional neutrino 
beam with a decay region to produce pion-decay-in-flight neutrinos for the MiniBooNE experiment and 
will run to support the MicroBooNE experiment.  Due to the short decay region, it also can serve as 
as a source of neutrinos from stopped pions in the target, horn, and surrounding structures.  It therefore 
could perform similar neutrino measurements as those outlined in the above section on the SNS.  
The maximum beam power the BNB is 32 kWatt, while the SNS is $\sim$1 MWatt.  
While the SNS has a factor of 30 higher beam power, it is possible the CAPTAIN detector at the BNB could 
be built as close as 10m to the absorber.  At the SNS, it is likely the detector would be at least 20 to 30 meters away from 
the target.  The BNB stopped pion flux at 10m from the absorber is estimated to be $2\times 10^{6} \nu/cm^2/s$, 
compared with $4.7\times 10^{6} \nu/cm^2/s$ \cite{BNBFlux} at 30m from the SNS target.   
The BNB may be a competitive choice for carrying out measurements of low-energy neutrinos.  
Further studies will be required to determine
neutron background rates and if it becomes a limiting factor in how close the detector can be to the BNB 
absorber.

\section{Other Neutrino Possibilities}

There are other neutrino running possibilities.  For example, space may be available in the SciBooNE hall 
in the Booster Neutrino Beamline at Fermilab.  While preliminary studies do not demonstrate a significant 
benefit to MicroBooNE from running a 5-ton near detector, the situation could change if any anomalies arise 
in MicroBooNE's first data.  

Other spallation neutron sources exist around the world.  If running at SNS or the BNB becomes problematic, 
there may be opportunities at other facilities.


%
%

\chapter{Conclusions}

\indent

The CAPTAIN program will make significant contributions to the development of LBNE.  
It will make unique physics measurements that support most important LBNE missions 
including: the long-baseline neutrino oscillation 
program, the atmospheric neutrino program (and correlated WIMP search) 
and the supernova neutrino program.  
The detector will also serve as a test bed for various calibration and detection strategy 
schemes - especially regarding laser calibration and photon detection.

The scientific program of CAPTAIN will make unique measurements of neutron 
interactions in argon as well as detailed neutrino interaction measurements.

\newpage


\begin{thebibliography}{99}
\bibitem{ref:WArP} R.~Acciarri {\it et al.} Effects of Nitrogen contamination in liquid Argon, 2010 JINST 5 P06003 (2009).
\bibitem{ref:FNAL-N2} B.~J.~P.~Jones {\it et al.} A Measurement of the Absorption of Liquid Argon Scintillation Light by Dissolved Nitrogen at the Part-Per-Million Level, arXiv:1306.4605 (2013).

\bibitem{ref:microboone} Chen, H. et al., {\emph{FERMILAB-PROPOSAL-0974} 2007}.

\bibitem{ref:lapd} B Rebel et al., {\emph{J. Conf. Ser.} {\bf 308}(2011) 012023}.

\bibitem{ref:conti} E. Conti et al., {\emph{Phys.\ Rev.\ B}{\bf 68}(2003) 054201}.

\bibitem{ref:doke} T. Doke et al., {\emph{Jpn.\ J.\ Appl.\ Phys.\ }{\bf 41}(2002) 1538}.

\bibitem{bibTPB}
V.M. Gehman et al., \emph{Fluorescence Efficiency and Visible
  Re-emission Spectrum of Tetraphenyl Butadiene Films at Extreme
  Ultraviolet Wavelengths},  {\emph{Nucl.\ Instr.\ Meth.\ A} {\bf 654} (2011) 116} [astroph{1104.3259v2}].

\bibitem{bibGastler}
D.E. Gastler, \emph{Design of Single Phase Liquid Argon Detectors for Dark Matter Searches}, Ph.D. Dissertation, Boston University, (2012).

\bibitem{ref:SunLaser}
J.~Sun {\it et al.} {\emph {Nucl.\ Instr.\ Meth.\ A} {\bf 370} (1996) 372}

\bibitem{ref:RossiLaser}
B.~Rossi {\it et al.} {\emph {JINST}{\bf 4}(2009)P07011}

\bibitem{bibKubota1}
S. Kubota, M. Hishida, S. Himi, M. Suzuki, \& J. Ruan, \emph{The suppression of the slow component in xenon-doped liquid argon scintillation}, {\emph{Nucl.\ Meth.\ Instr.\ A} {\bf 327} (1993) 71}.

\bibitem{bibKubota2}
M. Suzuki, M. Hishida, J. Ruan, \& S. Kubota, \emph{Light output and collected charge in xenon-doped liquid argon}, {\emph{Nucl.\ Meth.\ Instr.\ A} { \bf 327} (1993) 67}.

\bibitem{bibPollman}
P. Peiffer, T. Pollman, S. Schonert, A Smolnikov, \& S. Vasiliev, \emph{Pulse shape analysis of scintillation signals from pure and xenon-doped liquid argon for radioactive background identification}, {\emph{J.\ Instr.} {\bf 3} (2008) P08007}.

\bibitem{ref:Brooks}
M.~L.~Brooks {\it et al.} {\emph{Phys.\ Rev.\ C} {\bf 45} (1992) 2343}.

\bibitem{ref:argoneut}
C.~Anderson {\it et al.} {\emph{  arXiv:1111.0103 }}.

\bibitem{ref:forzel}
J.~A.~Formaggio and G.~P.~Zeller {\emph{ Rev.\ Mod.\ Phys.\ } {\bf 84} (2012) 1307 }. 

\bibitem{ref:duan2010}
H.~Duan, G.~M.~Fuller, and Y.~Quan {\emph{Ann.\ Rev.\ Nucl.\ Part.\ Sci.\ } {\bf 60} (2010) 569}.

\bibitem{ref:Bolozdynya2013}
A.~Bolozdynya {\it et al.}  arXiv:1211.5199.


\bibitem{BNBFlux}
Jonghee Yoo, \emph{Coherent Elastic Neutrino Nucleus Scattering (CENNS: P1040)}, 5 June 2013 Fermilab Physics Advisory Committee. 


\end{thebibliography}


\end{document}